# Approximations of Algorithmic and Structural Complexity Validate Cognitive-behavioural Experimental Results

A new approach to formally characterise Cognitive Decision Systems

**Hector Zenil**

Machine Learning Group,

Department of Chemical Engineering and Biotechnology,

University of Cambridge;

Kellogg College, University of Oxford &

Oxford Immune Algorithmics Ltd.

hector.zenil@cs.ox.ac.uk

**James A.R. Marshall**

Department of Computer Science,

Complex Systems Modelling research group,

University of Sheffield, United Kingdom

j.marshall@dcs.shef.ac.uk

&

**Jesper Tegnér**

Living Systems Laboratory,

Biological and Environmental Science and Engineering Division,

King Abdullah University of Science and Technology

jesper.tegner@kaust.edu.sa

# Abstract

Being able to objectively characterise the intrinsic complexity of behavioural patterns resulting from human or animal decisions is fundamental for deconvolving cognition and designing autonomous artificial intelligence systems. Yet complexity is difficult in practice, particularly when strings are short. By numerically approximating algorithmic (Kolmogorov) complexity ($K$), we establish an objective tool to characterise behavioural complexity. Next, we approximate structural (Bennett's Logical Depth) complexity ($LD$) to assess the amount of computation required for generating a behavioural string.

We apply our toolbox to three landmark studies of animal behaviour of increasing sophistication and degree of environmental influence, including studies of foraging communication by ants, flight patterns of fruit flies, and tactical deception and competition (e.g., predator-prey) strategies. We find that ants harness the environmental condition in their internal decision process, modulating their behavioural complexity accordingly. Our analysis of flight (fruit flies) invalidated the common hypothesis that animals navigating in an environment devoid of stimuli adopt a random strategy. Fruit flies exposed to a featureless environment deviated the most from Levy flight, suggesting an algorithmic bias in their attempt to devise a useful (navigation) strategy. Similarly, a logical depth analysis of rats revealed that the structural complexity of the rat always ends up matching the structural complexity of the competitor, with the rats' behaviour simulating algorithmic randomness.

Finally, we discuss how experiments on how humans perceive randomness suggest the existence of an algorithmic bias in our reasoning and decision processes, in line with our analysis of the animal experiments. This contrasts with the view of the mind as performing faulty computations when presented with randomised items. In summary, our formal toolbox objectively characterises external constraints on putative models of the "internal" decision process in humans and animals.

## Author Summary

*Behavioural sequences consist of a finite number of actions or decisions combined in various spatial and temporal patterns that can be analysed using mathematical tools. Understanding the mechanisms underlying complex human and animal behaviour is key to understanding biological and cognitive causality. In the past, mathematical tools for studying behavioural sequences were largely drawn from classical probability theory and traditional information theory. Our work here represents an advance in introducing powerful mathematical tools drawn from complexity science and information theory to study and quantify behavioural sequences' randomness, simplicity, and structure. In addition, the tools and concepts introduced here offer new and alternative means for behavioural analysis, interpretation, and hypothesis testing.*

# Introduction

To unravel the essence of the machinery enabling human and animal decisions is fundamental not only for understanding cognition but, by extension, when aiming to design advanced autonomous artificial intelligence systems. Recent progress in cognitive science suggests a Bayesian view of cognition as constituting a predictive system (Fahlman et al 1983; Rao and Ballard, 1999; Friston and Stephan, 2007; Friston, 2010). Such a view is succinctly captured in Shakespeare's in Macbeth, when Banquo asks the witches whether they can "*……… look into the seeds of time, and say which grain will grow and which will not*". At the core of this view is the notion that our cognitive apparatus incorporates a model of the world, serving as a basis for making decisions (Schmidhuber and Ha 2018; Friston et al, 2021). Thus, the Bayesian brain hypothesis suggests that underlying our behaviour there is an inherent cognitive structure for making decisions based on predictions, which are derived in turn from a real-world model.

Therefore in some sense, cognitive systems for decision making in animals and humans hinge upon the capacity to handle probabilities under a Bayesian model. Such probabilities are thought to be central to the decision process. Here we propose to reformulate the challenge of dealing with "internal" probabilities using the concept of complexity, since we then can objectively ask how random or non-random particular instances of observable "external" behaviour are. From such a viewpoint, we can potentially discover strong objective "external" regularisations constraining our models and analysis of the "internal" decision process, Bayesian or not. Since the emergence of the Bayesian predictive paradigm in cognitive science referred to above, researchers have expressed the need for a formal account of 'complexity' to, for example, enable objective characterisation of animal experiments. (e.g., Rushen, 1991; Manor 2013). There has, however, been a struggle to provide formal, normative, non-ad-hoc, and universal accounts of features in behavioural sequences that are more advanced than probabilistic tools.

Previous work has introduced some measures of *fractality* (spatial and temporal) and Shannon Entropy in applications of relevance to human and animal welfare (Sabatini, 2000; Costa et al., 2002; Manor 2012). Yet, little has been numerically attempted towards a systematic animal behavioural analysis using quantitative measures of algorithmic content. Relying primarily upon heuristic analysis of animal and human behavioural data, different ideas have been proposed regarding the nature of decisions, behaviour, probabilities, world models, and cognition which are thought to jointly account for behavioural data. For example, organisms exposed to a featureless environment have been thought to resort to a behavioural strategy close to a random walk, characterised by isotropic step lengths and following a heavy-tailed Levy distribution. Yet, without a proper formal analysis of "complexity," we risk misinterpreting behavioural data from animals and humans.

Here we build on earlier work (Gauvrit et al., 2014a; Gauvrit et al., 2014b; Gauvrit et al., 2015; Gauvrit et al., 2017a; Gauvrit et al., 2017b, Zenil et al., 2017, Zenil et al. 2020b) to show how algorithmic information theory provides measures for the high-order characterisation of processes produced by deterministic choices (Zenil et al., 2018; Zenil et al., 2019). Importantly, we target cases where such processes display no statistical regularities or rankings of ordered versus random-looking sequences in terms of their information content. Our objective in responding to this challenge is to use algorithmic information theory to develop a quantitative toolbox for complexity, and to validate our tools using well-known landmark studies of animal behaviour. This task entails developing tools that can handle limited amounts of data in the behavioural sciences, i.e., estimating complexity for short strings. One problem for a decision system is how to build a prior under several partly competing constraints. For example, how do we meet the criteria for neutrality, making a minimal number of assumptions (Ockham's razor), while also taking into account all possible scenarios (Epicurus' Principle of Multiple Explanations) and moreover being sufficiently informative beyond statistical uniformity (Gauvrit et al., 2015).

To this end, we introduce Kolmogorov complexity (*K*) as a coherent formalism, which has been developed to incorporate all these principles. Kolmogorov complexity quantifies simplicity versus randomness and enables a distinction between correlation and causation in data. Its Kolmogorov complexity quantifies the complexity of an object by the length of its shortest possible description. For example, low *K* means that digits in a sequence can be deterministically generated--- each bit is causally connected by a common generating source. However, that a sequence has a large Shannon Entropy means that its digits do not look statistically similar, not that they are necessarily causally disconnected. Thus far, statistical lossless compression algorithms have been used to approximate Kolmogorov complexity. However, they are based on statistical properties of sequences that are used to encode regularities, and as such they are, in fact, Shannon Entropy rate estimators (Zenil, 2020a). Here we use an alternative method to approximate *K* based on Algorithmic Probability (Zenil & Delahaye, 2011; Soler-Toscano et al., 2014). In contrast to lossless compression, we can also deal with short sequences typically encountered in the behavioural sciences.

In addition, we also target what is referred to as the structural complexity, i.e., the amount of computation performed by a human or animal in decision-making. Since such a distinction between structural complexity and randomness cannot be accomplished using Kolmogorov complexity, we make use of the notion of Logical Depth (*LD*) (Bennett, 1998). Importantly, our approximation technique for short sequences also allow an estimation of *LD* based upon and motivated by Logical Depth (Zenil et al. 2012c; Zenil et al., 2020b).

## Methods

The algorithmic information theory provides a formal framework for verifying intuitive notions of complexity. For example, the algorithmic probability of a sequence *s* is the probability that a randomly chosen program running on a Universal (prefix-free) Turing Machine will produce *s* and halt. Since Turing Machines are conjectured to be able to perform any algorithmically specified computation, this corresponds to the probability that a random computation will produce s. It therefore serves as a natural formal definition of a prior distribution for Bayesian applications. Also, as we will see in the next section, the algorithmic probability of a string *s* is negatively linked to its (Kolmogorov-Chaitin) algorithmic complexity, defined as the length of the shortest program that produces *s* and then halts (Kolmogorov 1965; Chaitin 1969).

One important drawback of algorithmic complexity (which we will denote by *K*) is that it is not computable. Or, more precisely, *K* is lower-semi-computable, meaning that it cannot be computed with absolute certainty but can be arbitrarily approximated from above. Indeed, statistical lossless compression algorithms have been used to approximate *K*. However, new methods more deeply rooted in and motivated by algorithmic information theory can potentially provide alternative estimations of algorithmic probability (Delahaye and Zenil, 2011), and thus *K*, of strings of any length, especially short ones (c.f. next section), which has spurred a renewed interest in exact and objective numerical approximations of behavioural sequences. One feature that has been the basis of a criticism of applications of $K$ is that $K$ assigns the greatest complexity to random sequences. The notion of "sophistication" is useful to broaden the analysis of behavioural sequences. It assesses the structure of the sequence and assigns low complexity to randomness, since the measure introduces computational time. One such measure derives from Charles Bennett's seminal contribution, based on Kolmogorov complexity, and is referred to as Logical Depth. Here we also use a measure based upon or motivated by Logical Depth (Zenil et al. 2012c) to quantify the complexity of behavioural sequences because it can provide further insight into another important aspect of animal behaviour beyond randomness, namely, the "computational effort" that an animal may invest in behaving in a particular way.

## Algorithmic probability

One long-standing and widely used method for assessing Kolmogorov-Chaitin complexity is lossless compression, epitomised by the Lempel-Ziv algorithm. This tool, together with classical Shannon Entropy (Wang et al., 2014), has been used recently in neuropsychology to investigate the complexity of electroencephalogram data (EEG) or Functional Magnetic Resonance Imaging (fMRI) data (Maguire et al. 2014; Adenauer et al., 2013). The size of a compressed file indicates its algorithmic complexity, and the size is, in fact, an upper bound of the true algorithmic complexity. However, compression methods have a basic flaw: they can only recognise statistical regularities, and are therefore implementations of variations of entropic measures, only assessing the rate of entropic convergence based on repetitions of strings of fixed sliding-window size. If statistical lossless compression algorithms work for approximating Kolmogorov complexity, they do so because compression is a sufficient test for non-randomness. Yet, statistical lossless compression fails in the other direction, unable to tell whether something is the result of or is produced by an algorithmic process (such as the digits of the mathematical constant π). That is, they cannot detect structure outside of simple repetition.

Popular compression methods are also inappropriate for short strings (of, say, less than a few hundred symbols). For short strings, lossless compression algorithms often yield files that are longer than the strings themselves, providing very unstable results that are difficult, if not impossible, to interpret (see Table 1). In cognitive and behavioural science, however, researchers usually deal with short strings of a few tens of symbols, for which compression methods are useless. This is one of the reasons behavioural scientists have long relied on tailor-made measures of complexity instead, because the factors mentioned above limit the applicability of lossless compression techniques for approximating complexity in behavioural sequences.

The *Coding theorem method* (Delahaye and Zenil, 2012; Soler-Toscano et al., 2013) has been specifically designed to address this challenge. Thanks to this method, researchers have defined "Algorithmic Complexity for Short Strings" (ACSS). This is a concrete approximation of algorithmic complexity (Gauvrit et al., 2015), usable with very short strings. ACSS is available freely as an R-package (Gauvrit et al., 2015) and through an online complexity calculator (http://www.complexity-calculator.com).

The idea at the root of ACSS is the use of algorithmic probability as a means to capture algorithmic complexity. The algorithmic probability of a string *s* is defined as the probability that a universal prefix-free reference Turing machine *U* will produce *s* and halt. Formally (Levin, 1974),

$$(1) \quad m(s) = \sum_{U(p) = s} 1/2^{-|p|}$$

The algorithmic complexity of a string s is defined as the length of the shortest program *p* that, running on a reference universal prefix-free Turing machine *U*, will produce s and then halt. Formally (Kolmogorov, 1965; Chaitin, 1969),

$$(2)\quad K(s) = min\{|p|, U(p) = s\}$$

K(s) and m(s) both depend on the choice of the Turing machine *U*. Thus, the expression "the algorithmic complexity of *s*" is, in itself, a shortcut. For long strings, this dependency is relatively small. Indeed, the invariance theorem states that for any two universal prefix-free Turing machines *U* and *U'*, there exists a constant *c* independent of s such that (Solomonoff, 1964; Kolmogorov, 1965; Chaitin, 1969)

$$(3)\quad | K_U(s) – K_{U'}(s) | < c$$

The constant *c* can be arbitrarily large. If one would like to approximate the algorithmic complexity of short strings, the choice of *U* is thus relevant.

To partially overcome this inconvenience, we can take advantage of a formal link established between algorithmic probability and algorithmic complexity. The algorithmic coding theorem reads (Levin, 1974)

$$(4)\quad K_U(s) = – log_2(m_U(s)) + O(1)$$

This theorem can be used to approximate a $K_U$ *(s)* through an estimation of $m_U(s)$, *where* $K_U$ and $m_U$ are approximations of *K* and *m* obtained by using a "reference" universal Turing machine *U*. Instead of choosing a particular "reference" universal Turing machine and feeding it with programs, Delahaye and Zenil (2007 and 2012) used a huge sample of Turing machines running on blank tapes. By doing so, they built an experimental distribution approaching *m*(s) more smoothly, averaging over many Turing machines. ACSS(s) was then defined as $–log_2(m(s))$ *by the algorithmic Coding theorem (eq. 4)*. ACSS(s) approximates an average $K_U(s)$. To validate the method, researchers have studied how ACSS varies under different conditions. It has been found that ACSS, as computed with different samples of small Turing machines, remains stable (Gauvrit et al., 2014). Also, several descriptions of Turing machines did not alter ACSS (Zenil et al., 2012). For example, Zenil et al. (2012) showed that ACSS remained relatively stable when using cellular automata instead of Turing machines. On a more practical level, ACSS is also validated by experimental results. For instance, as we will see in the following sections, ACSS is linked to human complexity perception. Moreover, the transition between using ACSS and lossless compression algorithms is smooth, with the two behaving similarly when the scope of string lengths overlap.

## Bennett's Logical Depth

As noted in the Introduction, a measure of the "sophistication" of a sequence can be arrived at by combining the notions of algorithmic complexity and computational time. According to the concept of Logical Depth (Bennett, 1988), the complexity of a string is best defined by the time that an unfolding process takes to reproduce the string from its shortest description. The longer it takes, the more complex the string. Hence, complex objects can be seen as "containing internal evidence of a nontrivial causal history" (Bennett, 1998).

Unlike algorithmic complexity, which assigns a high complexity to random and highly organised objects, placing them at the same level, Bennett's Logical Depth assigns a low complexity to both random and trivial objects. It is thus more in keeping with our intuitive sense of the complexity of physical objects, because trivial and random objects are intuitively easy to produce, lack a lengthy history, and unfold quickly. Bennett's main motivation was to provide a reasonable means for measuring the physical complexity of real-world objects. Bennett provides a careful development (Bennett 1988) of the notion of logical depth, taking into account the near-shortest programs, not merely the shortest one, to arrive at a robust measure. For finite strings, one of Bennett's formal approaches to the logical depth of a string is defined as follows:

Let *s* be a string and *d* a significance parameter. A string's depth at some significance *d*, is given by

$$(5)\quad LD_d(s)=min\{T(p) : (|p| - |p'| < d) \text{ and } (U(p) = s)\},$$

where $T(p)$ is the number of steps in the computation $U(p) = s$, and $|p'|$ is the length of the shortest program for *s*, (thus $|p'|$ is the Kolmogorov complexity $K(s)$). In other words, $LD_d(s)$ is the least time *T* required to compute s from a *d*-incompressible program p on a Turing machine *U*, that is, a program that cannot be compressed by more than a fixed (small) number of bits *d (*Bennett, 1998). For algorithmic complexity, the choice of a universal Turing machine is bounded by an additive constant (as shown by the Invariance theorem described in the previous section). In contrast, Logical Depth is bounded by a multiplicative factor (Bennett 1989). The simplicity of Bennett's first definition of Logical Depth (Bennett, 1988), independent of *size significance*, makes it more suitable for applications (Zenil and Delahaye, 2011), serving as a practical approximation to this measure via the decompression times of compressed strings. To this end, it uses lossless compression algorithms, whose deviation from perfect compression is unknown (and cannot generally be known due to uncomputability results), to calculate size significance. This is because where LD is concerned, it is more relevant to consider the shortest time vis-à-vis a set of near-smallest programs rather than just a single, perhaps unrepresentative, time required by the shortest program alone. We will denote by $LD(s)$ a measure approximating the Logical Depth of a string s, with no recourse to the significance parameter, and approximated by a powerful method, an alternative to lossless compression, explained in the next section.

## Numerical estimation of algorithmic probability and *K*

We use the concept of algorithmic probability for the calculation of *K* (and *LD*) by application of the algorithmic Coding theorem. First, a sample of $2{,}836 \cdot 10^9$ random Turing machines are selected from a reduced enumeration of all 5-state 2-symbol Turing machines (Soler-Toscano et al., 2013), using the standard Turing machine formalism of the 'Busy Beaver' problem (Rado, 1962; Brady, 1983). The sample output returns the string produced by the halting machines, their runtimes, and the instructions used by each Turing machine. All the necessary information to estimate both the algorithmic probability measure and Logical Depth by finding the smallest machine (for this Turing machine formalism) producing *s* before halting is presented in Figure 1.

## Block Decomposition Method

The Block Decomposition Method (Zenil et al. 2014) is used to estimate an algorithmic probability of longer sequences based on the concept of algorithmic probability, the calculation of which is computationally infeasible, given the number of Turing machines that it would be necessary to run in order to have a statistically significant frequency value, and ultimately impossible because of incomputability results. However, unlike traditional statistical lossless compression algorithms used to estimate *K*, the Block Decomposition Method (BDM) can deal with short sequences and has been successfully applied to human cognition before (Gauvrit et al., 2014b and Kempe et al., 2015b), as well as to algebraic and topological characterisations of graph theory and complex networks (Zenil et al., 2014), and offers an alternative to the use of statistical measures, including popular compression algorithms such as LZW that are mostly entropy estimators (Zenil, 2020a).

The method consists in decomposing a long sequence into shorter, optionally overlapping sub-sequences, the Kolmogorov complexity of which can be estimated by running a large set of random computer programs in order to estimate their algorithmic probability. For example, the sequence “123123456” can yield, with a 6-symbol window with a 3-symbol overlap, the two subsequences “123123” and “123456”, i.e., what is known as *k-mers*, in this case 6-mers, all substrings of length 6. BDM then takes advantage of possible repetitions by applying the following formula:

$$(6) \quad C(s)=\sum_p(\log_2(n_p)+K(p)),$$

where the sum ranges over the different subsequences $p$, $n_p$ is the number of occurrences of each $p$, and K($p$) is the complexity of $p$ as approximated by CTM implemented in ACSS. As the formula shows, the Block decomposition method (BDM) takes into account both the local complexity of the subsequence and remote regularities by repetition of the same substring in the original sequence.

BDM is a greedy divide-and-conquer algorithm that extends the power of the algorithmic Coding Theorem Method (see Figure 1), which is computationally intractable in a sound information-theoretical fashion, given classical information theory. We know we can encode the complexity of an object repeated $n$ times with the complexity of the object and only $\log_2(n)$ bits for the number of repetitions, thus helping to find tighter upper bounds to $K$. Unlike lossless compression algorithms used to approximate $K$, CTM, with BDM, constitutes a more deeply-rooted algorithmic approach to $K$, rather than merely statistical ones (such as LZW and cognates), which are more suitable for estimating *Entropy rate* than algorithmic (Kolmogorov) complexity.

This is because, unlike statistical lossless compression (e.g. LZW), CTM can find the computer programs that produce certain non-statistical algorithmic patterns, such as subsequences of the decimal expansion of the mathematical constant π, that will merely look random to a popular statistical lossless compression algorithm.

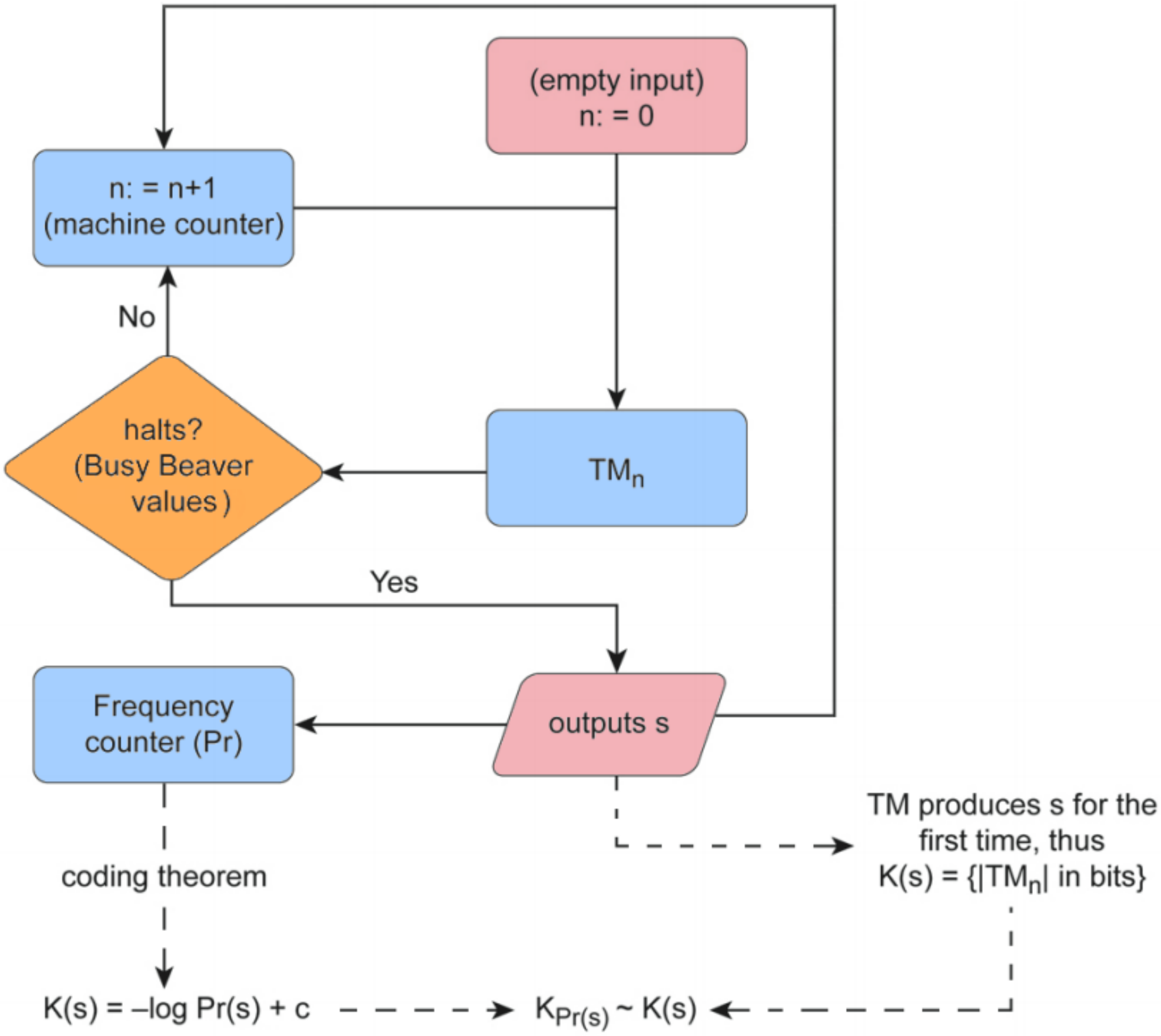

**Figure 1. Flow chart of the Coding Theorem Method (CTM) and its implementation Algorithmic Complexity for Short Strings (ACSS), available as a package for the R programming language, shows how *K* and *LD* are approximated by way of an Algorithmic Probability-based measure upon application of the algorithmic Coding theorem. The method consists in running an enumeration of Turing machines from smaller to larger (by number of states) and determining whether the machine halts or not for a large set of small Turing machines for which either their halting time can be decided (with the so-called Busy Beaver functions) or an educated theoretical and numerical guess made. The algorithm counts the number of times a sequence is produced, which is a lower bound on its algorithmic probability, and then is transformed to an upper bound of the sequence's Kolmogorov complexity using the so-called algorithmic Coding theorem (bottom left). The same algorithm also finds the shortest, fastest Turing machine that produces the same sequence, hence providing an estimation of its Logical Depth (bottom right). An online tool that provides estimations of CTM, and soon will include BDM too, can be found at http://www.complexity-calculator.com**

## Application to animal behavioural data

To test the usefulness of algorithmic probability and logical depth for capturing intuitive notions of behavioural complexity, we applied these techniques to several experimental datasets for which complexity arguments are either lacking or have only been informally advanced. These datasets come from studies of the foraging behaviour of *Formica* ants in binary mazes (*e.g.* Reznikova, 1996), and of the flight behaviour of *Drosophila* flies in virtual arenas (*e.g.* Maye et al., 2007).

### Ant foraging and the complexity of instructions in binary mazes

In a series of experiments surveyed in Reznikova(1996), red-blood ant scouts (*Formica sanguinea*) and red wood ants (*Formica rufa*) were placed in a binary maze with food (sugar syrup) at randomly selected endpoints. Scout ants were allowed to return through the maze once they had found the food to communicate instructions to a foraging team from the colony; scout ants were claimed to translate sensorimotor experiences into numerical or logical prescriptions transmitted to conspecifics in the form of sequences of branches (left or right) to follow to find the food successfully. 335 scout ants and their foraging teams took part in all the experiments with the binary tree mazes, and each scout took part in ten or more trials. 338 trials were carried out using mazes with 2, 3, 4, 5, and 6 forks. The scout ants were observed to take progressively longer to communicate paths in deeper mazes (with more turns); that is, they transmitted more information. Informal results suggested that algorithmically simpler instructions of “right” and “left” movements toward the food patches were communicated faster and more effectively than more Kolmogorov complex (random-looking) behavioural sequences, thus suggesting that when strings are of the same length but transmitted at different rates, ants are capable of compressing the simpler sequences, unlike the more complex ones that are harder to compress or are uncompressible. This was not formalised, however. When conducting the ant experiment, the researchers also found that almost all naive foragers were able to find food on their own but that the time they spent was 10-15 times longer than the time spent by those ants that entered the maze after contact with a successful scout bearing information about the location of the food (Ryabko and Reznikova, 2009). Reznikova et al. (Ryabko and Reznikova, 2009; Reznikova and Ryabko, 2011; Reznikova and Ryabko, 2012) and other researchers (Li and Vitanyi, 2008) were not able to numerically validate the relationship between complexity and communication times suggested by the results of the ant experiment. We applied our techniques for approximating algorithmic probability and logical depth (Delahaye & Zenil, 2012; Soler-Toscano et al., 2013; Zenil et al., 2020b) to these extant data.

## Low random behaviour of fruit flies in the absence of stimuli

Experiments with *Drosophila* examined the behaviour of tethered flies in a flight simulator consisting of a cylindrical arena homogeneously illuminated from behind (Maye et al., 2007). A fly's tendency to perform left or right turns (yaw torque) was measured continuously and fed into a computer. The flies were divided into three groups: the 'openloop' group flew in a completely featureless white panorama (*i.e.*, without feedback from the uniform environment–open loop). In addition to the open-loop group, data from two control groups were analysed. These groups flew in an arena with either a single stripe ('onestripe' group) or in a uniformly dashed arena ('uniform' group). The 'onestripe' group's environment contained a single black stripe as a visual landmark (pattern) that allowed for straight flight in closed-loop control since the fly could translate its visual input into a yaw torque to control the angular position of the stripe. The 'uniform' group flew in a uniformly textured environment otherwise free of singularities. This environment was closed-loop in the same sense as that provided for the 'onestrip' group, since the fly could use its yaw torque to control the angular position of the uniform textured environment. Maye *et al.* (2007) concluded that in the featureless environment, fly behaviour was non-random, with the distribution of yaw directions produced by flies in the 'open-loop' group significantly deviating from the null Poisson distribution.

## Animal behaviour in environments of increasing complexity

A virtual competitive setting was designed as described in Tervo et al. (2014). Here, the algorithm played the role of a virtual competitor (that can also be seen as a predator) against a rat. The algorithm was programmed to predict which hole the rat would choose against three increasingly complex predictive competitors. The rat had to choose a hole that it thought would not be chosen by the competitor in order to be rewarded. The task is therefore a prediction task, where the more successful the rat is at predicting the competitor's behaviour, the better it can avoid it and be rewarded. The first competitor consisted of an algorithm based on a binomial test able to react to significant bias. For example, if a rat had a clear preference for choosing either the left or right hole, Competitor 1 would correspondingly predict the hole with statistical bias. Competitor 2 reacted to any bias based on a binomial test. Competitor 3 , however, displayed a diverse range of features and therefore constituted a greater challenge to the rat, because rats were rewarded with food if their competitors did not predict their hole choice. Each "environment" consisted of a long sequential list of trials for every competitor against 12 fresh individuals (rats). Competitors 1 and 2 used conditional prevalence of the left and right choices, given a particular history pattern of up to three prior steps, to inform their prediction. The optimal deterministic strategy consisted thus in keeping track of every pattern up to that length and ensuring that the conditional prevalence of going left or right was 0.5 (Tervo et al., 2014). The authors quantified how different the observed behaviour was from this optimal strategy by calculating the Kullback-Leibler divergence from the

optimal---a variation of Shannon Entropy---of the observed distribution of conditional prevalences, given all small sequences of lengths n = 1, 2, and 3. Their results, backed by clinical results tracking the engagement of the anterior cingulate cortex, an area of the brain related to decision-making, showed that when rats are faced with competitors that they cannot outsmart, they switch from strategic counter prediction behaviour to a stochastic mode.

Competitor 1 only looked for biases toward Left or Right. For example, if the rat had chosen the Left hole with probability 0.6, then Competitor 1 would have predicted that the rat would choose Left, which would have resulted in no reward for the rat. The rat would then switch to more complicated strategies, maximising Left and Right probability near 0.5 but following a deterministic bias in favour of itself. Thus, these animals were eventually able to model aspects of the underlying prediction algorithm and used that knowledge to anticipate the opponent's predictions. Virtual Competitor 1 did not display the brain's full capacity for generating behavioural variability, as it represented no challenge to the animals. Competitor 2 used the same prediction algorithm as Competitor 1, except that it removed the requirement that the bias in favour of one side or the other reach a predetermined threshold before competitive pressure was applied (see details in Tervo et al., 2014) and was therefore slightly more challenging for the rats. Competitor 3, however, used a more sophisticated machine-learning strategy that learned to generate a strong prediction based on a set of weak trends in the data (details are given in Friedman et al., 2000). The rationale of the original experiment was that a stronger competitor would detect some of the patterns that a weaker competitor missed, leading to a correct prediction of the choices made by the animal---and thus to withholding of the reward. So, Competitor 2 lookedfor biases on moves up to 3 times back, which maked it more complicated. Competitor 3 used a much more sophisticated learning algorithm. As can be seen from the plots, the complexity of the reward (prediction) is almost equal and, in some cases, greater than the complexity of the animal choices, unlike the cases involving Competitors 1 and 2, where the reward is not only greater but diverges asymptotically from the rat's complexity, indicating that the rat found an optimal behavioural strategy tending toward infinite reward vis-à-vis these competitors' learning strategies. The rat seemed to have a short learning period which it spent matching the complexity of the competitor's behaviour before finding out that the competitor was highly predictable. This behaviour allowed the rat to reduce its complexity and receive increasing rewards for lower behavioural complexity by fooling the competitor, e.g., with simple repetitions of LRR or RLL where the competitor expected L and R respectively as the last choice after LR and RL.

# Results

## Validation of numerical estimations of *K* and *LD* by numerical correlation

Here, we compare different ways of estimating *LD* and K by running a large set of enumerated standard Turing machines of increasing size size (Zenil et al., 2020b). We computed the correlation between approximations to K and LD utilising lossless compression and CTM for all $2^{12}$ bit strings up to length 12. To quantify the analysis, we use a nonlinear least-squares regression model. We used the computed data and curve fitting of Compression and CTM estimations of K against CTM estimations of LD. Here we find the best fit using a quadratic function $14.76 + 0.0077x - 6.4 \cdot 10^{-7}x^2$. We find that CTM displays lower LD values for low and high K values. This demonstrates that only CTM conforms to the theoretical expectation where both Kolmogorov simple and Kolmogorov random strings are assigned lower Logical Depth. In contrast, lossless compression, instantiated using both Compress and BZip2 algorithms, displayed a trivial correlation with LD. This demonstrates that approximating *K* by traditional lossless compression algorithms, such as LZW and cognates, does not conform to the theoretical expectation of low LD for the lowest and highest *K* values. Moreover, nor can they be usefully applied to short behavioural sequences (Delahaye and Zenil, 2011 and Soler-Toscano et al., 2014), such as the ones from a well-known behavioural experiment with ants (Table 1) and in more recent studies of the behaviour of fruit flies and rats.  This is because random sequences, like trivial ones, do not result from a sophisticated computation of their shortest descriptions. Indeed, if a sequence is random, its shortest description is of about the same length as the sequence itself, and therefore the decompression instructions are very short or non-existent. And if a sequence is trivially compressible, then the instructions for decompressing it may also be expected to be very simple. Right between these extremes, we find high sophistication or logical depth.

## *K* and *LD* estimations of animal behavioural sequences

Here we demonstrate, in three different case-studies, that applying the algorithmic Coding theorem and numerical estimations based upon or motivated by Algorithmic Probability and Logical Depth, provides objective quantification of animal behavioural sequences (Zenil et al., 2020b).

### Ants' communication times follow complexity of foraging instructions

Interestingly, the order in which Reznikova et al. (Ryabko and Reznikova, 1996) informally sorted ants' behavioural sequences, it turns out, corresponds to an order of increasing Kolmogorov complexity, as approximated by our methods motivated by and based upon algorithmic probability (Zenil et al., 2020b). Table 1 shows the main results, comparing sequence complexity and scout-forager communication times.

**Table 1. Movement and duration (as reported by Ryabko and Reznikova, 1996) of information transmission from F. sanguinea scouts to foragers; 0 encodes a right turn, and 1 encodes a left turn in a binary maze. E(s) is the Shannon Entropy of s. In contrast, CTM(s), C(s), and Bzip2(s) are all approximations of K(s), with CTM(s) the approximation of K(s) by Algorithmic Probability and C(s) and Bzip2(s) by lossless compression (Compress and BZip2). LD(s) is the approximation of Logical Depth by CTM. The table is sorted by duration, and it is clear that neither Entropy nor lossless compression (E, C, and Bzip2) offers enough resolution to separate the complexity of behavioural sequences into more than 2 groups because E counts the number of different symbols. Compression is a variation of Entropy rate but also fails at compressing short strings. However, CTM and LD are highly correlated, showing an increase in complexity, as suggested by the authors of the original experiments.**

| *Ant behavioural sequence (turn pattern)* | *Duration (mean sec)* | *E(s) (bits)* | *CTM(s) (bits)* | *C(s) (bytes )* | *Bzip2(s) (bytes)* | *LD(s) (number of instructions)* |
|---|---|---|---|---|---|---|
| 110 | 69 | 0.636 | 6 | 64 | 37 | 3 |
| 11 | 72 | 0 | 3 | 64 | 37 | 2 |
| 000 | 75 | 0 | 4 | 64 | 37 | 3 |
| 111 | 84 | 0 | 4 | 64 | 37 | 3 |
| 00000 | 78 | 0 | 4 | 64 | 39 | 5 |
| 000000 | 88 | 0 | 5 | 64 | 39 | 7 |
| 111111 | 90 | 0 | 5 | 64 | 39 | 7 |
| 1011 | 100 | 0.562 | 6 | 64 | 37 | 4 |
| 0110 | 120 | 0.693 | 7 | 64 | 38 | 4 |
| 101010 | 130 | 0.693 | 8 | 64 | 37 | 7 |
| 010101 | 135 | 0.693 | 8 | 64 | 37 | 7 |
| 00101 | 150 | 0.673 | 7 | 72 | 37 | 5 |

| 010001 | 180 | 0.63<br>6 | 8 | 72 | 38 | 7 |
|---|---|---|---|---|---|---|
| 101101 | 200 | 0.63<br>6 | 8 | 64 | 38 | 7 |
| 001000 | 220 | 0.45 | 7 | 72 | 37 | 7 |

The (Pearson) correlations found among the values reported in Table 1 and plotted in Figure 2 are as shown in Table 2.

**Table 2. CTM and Logical Depth found the best correlation—with high significance—between duration and Kolmogorov complexity (Figure 2). This formalises and provides a theoretical justification of the previous results, which were solely based upon visual inspection and intuition.**

| ***Index*** | ***Correlation*** | ***p-value*** |
|---|---|---|
| Entropy | 0.616 | 0.014 |
| Compress (LZ) | 0.61 | 0.014 |
| Bzip2 | 0 | 1 |
| CTM | 0.82 | 0.00017 |
| LD | 0.75 | 0.0013 |

More precisely, ant communication is correlated to the complexity of the instructions, i.e., instructions that take less time to communicate have low Kolmogorov complexity, while less efficiently communicated instructions have higher K and higher LD.

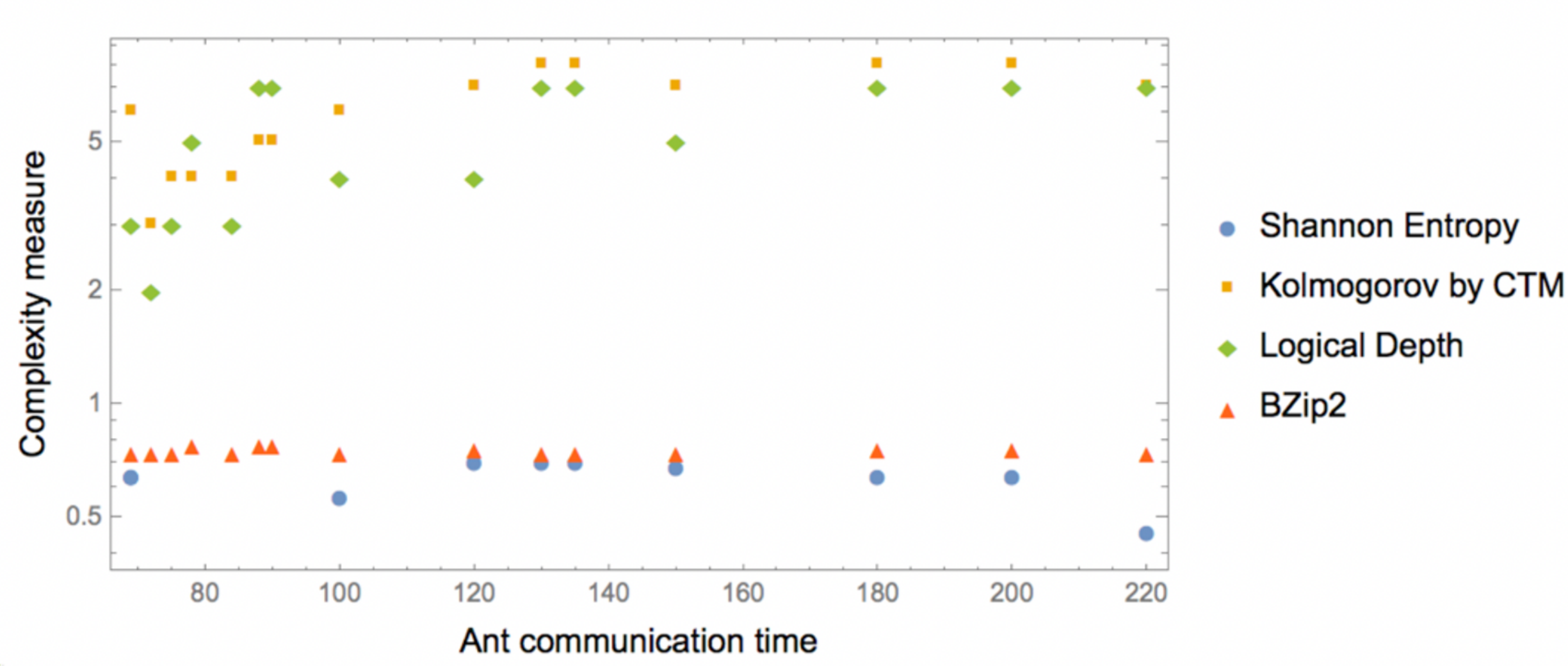

**Figure 2. Log correlation between normalised (by a common constant) complexity measures and ant communication time. Kolmogorov complexity (employing an Algorithmic Probability based CTM measure) and Logical Depth (also measured by BDM) display the greatest correlation, with longer times for more complex K and LD. Shannon Entropy and lossless compression display no sensitivity to differences in the sequence. Compress behaves like Entropy.**

### Fruit flies' display non-random behaviour in environments with little or no input stimuli

In the *Drosophila* experiments (Maye et al., 2007), yaw torque binary (right or left directions) behavioural sequences for tethered flies in a virtual reality flight arena were recorded in three environments for up to 30 minutes (see Methods).

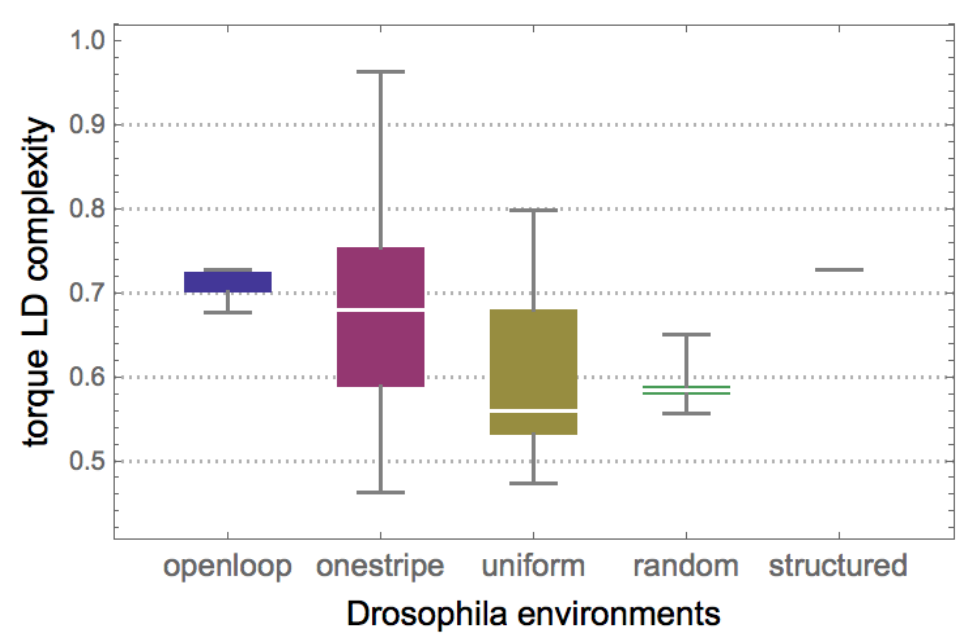

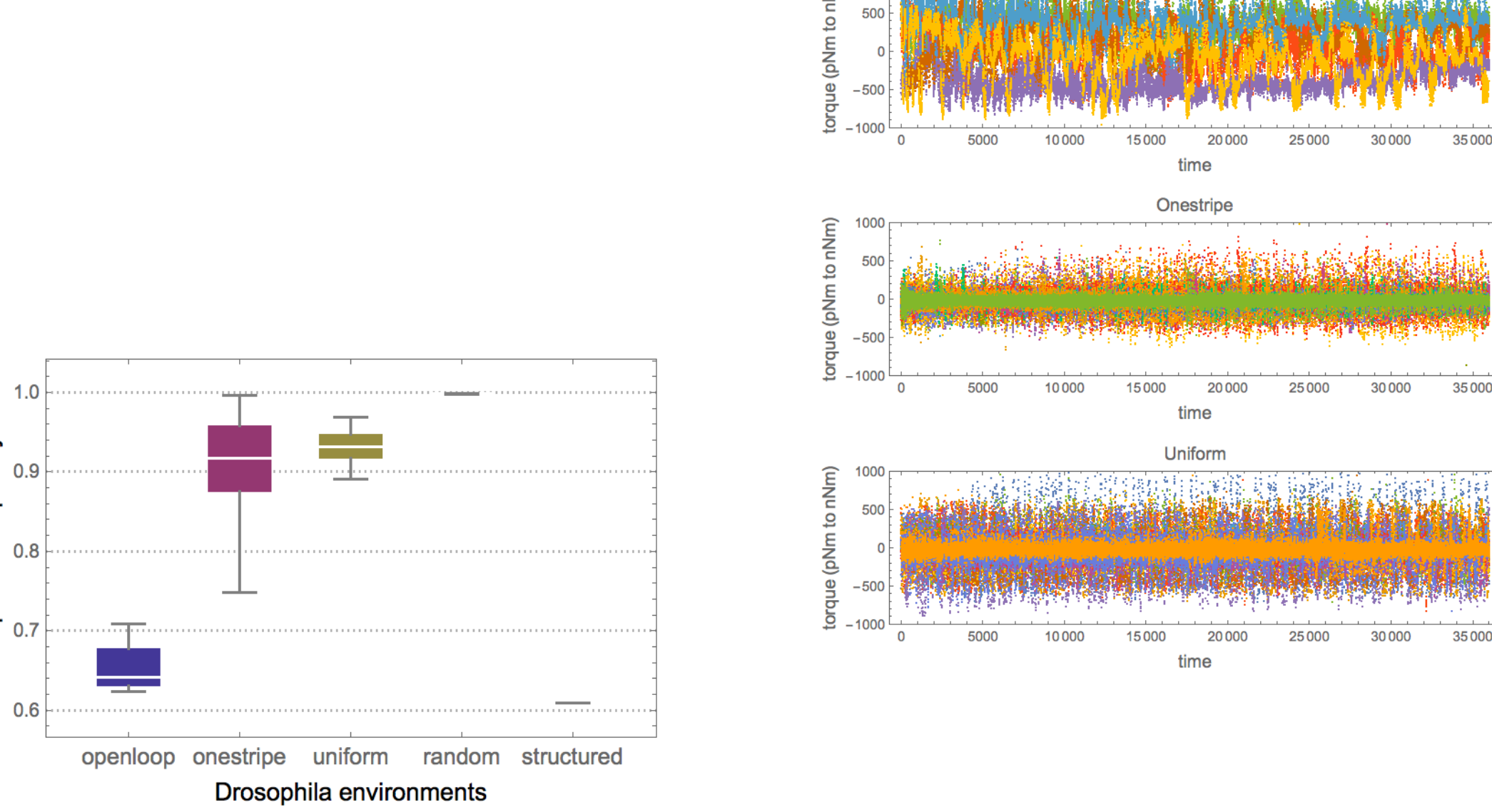

**Figure 3 Top left: Raw yaw torque series of the three fruit fly groups (about 38,000 data points from a 30-minute recorded flying period). Right top: Box plot of the series Kolmogorov complexity as measured by CTM over all flies in the three groups (6, 18, and 13). Each replicate was given a different colour. Right: In agreement with the results of Maye et al. (2007) the median complexity of the open-loop group is the most removed from (algorithmic) randomness. The values were normalised by maximum K complexity according to CTM (as compared to the substrings of greatest complexity as calculated from CTM). Bottom left: Box plot of the series' Logical Depth complexity as measured by CTM over all flies in the three groups. The results suggest that the open-loop group has the highest significant median structural complexity (Logical Depth), hence suggesting greater causal history or calculation and a greater remove from randomness. In contrast, the other two groups are closer to pseudo-randomly generated sequences using a log uniform PNRG. This supports the original authors' findings but adds that uniform stimuli seem to have a lower effect than no stimulus, and that the absence of stimuli also leads to different behaviour, perhaps as a strategy to elicit environmental feedback. In both plots, all values were normalised by maximum LD complexity to have them between 0 and 1. In both box plots, the trivial sequences consist of 1s, and therefore are maximally removed from randomness for Kolmogorov complexity (but are closer to randomness in the LD plot).**

As shown in Figure 3, the series of fruit fly torque spikes for the three groups of fruit flies had different Kolmogorov complexities, with the open-loop group being the furthest removed from algorithmic randomness and high in logical depth, suggesting an algorithmic source. This strengthens the authors' (Maye et al., 2007) conclusion that the open-loop group had the largest distance to randomness. It suggests that fly brains are more than just input/output systems and that the uniform group came closest to a characteristic Levy flight. This observation amounts to a falsification of the alternative hypothesis that flies behave randomly in the absence of stimuli, as their neurons would only fire erratically, displaying no pattern. Here, however, the results suggest that in the absence of stimuli, the flies are even more challenged to find different flight strategies, perhaps seeking stimuli that would provide feedback on their whereabouts. Moreover, as shown in Figure 3, by measuring the subtle differences in complexity along the various sequences, the variance provides another dimension for analysis---of how different flies reacted to the same environment. As shown by (Zenil et al., 2013b), if the environment is predictable, the cost of information processing is low. Still, this result also suggests that if the environment is not predictable, the cost of information processing quantified by logical depth is low if the stochastic behaviour strategy is adopted, providing an evolutionary advantage (Zenil et al., 2013b). This strengthens their conclusion that environmental complexity drives the organism's biological and cognitive complexity.

**Rats switching to random behaviour in environments of increasing complexity**

To the experimental data of Tervo et al. (2014) (binary behavioural sequences representing L for Left and R for right depending on the choice of the animal and the prediction of a virtual competitor), we applied lossless compression, BDM and BDM LD. BDM and compression showed the expected complexity for each set environment (see Figures 4 and 5). The first one (Competitor 1) displays the lowest randomness for both the animal and the competitor/reward sequences (the competitor's behaviour is a Boolean function of the reward sequence given by (choice & reward) or (~choice & ~reward)). The reward sequence encodes whether or not the competitor predicted the animal's choice and whether it was given a droplet---when avoiding the competitor's choice---or not. The goal of the rats is to outsmart the competitor's behaviour. Yet, that does not necessarily mean an increase in a rat's behavioural Kolmogorov complexity (randomness), because it can fool the competitor with a simple strategy, as is the case for Competitors 1 and 2, and by switching to random-looking behaviour in the case of Competitor 3. This variation of animal complexity over time, going from high to lower complexity, can be observed in Figure 4 and in Table 3, where a ranking of learning capabilities per rat for every environment is reported. In the environment with Competitor 1, a simple 3-tuple behavioural strategy outsmarts the competitor's behaviour with a strategy of low complexity by keeping the frequency of R and L choices at about 0.5 but following elementary patterns that the competitor does not follow (the alternation of LLR and RRL or RLL and LRR). On the other hand, logical depth shows that the structural complexity of the animal always ends up matching the structural complexity of the competitor and is never more or less than is required to outsmart it. From BDM and compression, it is also clear that the 3rd environment is of lower structural complexity. It's consistent with the conclusion of the clinical experiment that rats switched to stochastic behaviour when they could not outsmart the competitor, even though the mean is very similar when BDM and Compress are of lower complexity. This is not a surprise because the behaviour of the competitor is not algorithmically random; it only appears random to the rat, which in turn tries to behave randomly. But it does so not with greater complexity than the competitor but only by matching the complexity of the competitor. After all, it is unlikely that the rat is behaving in an algorithmically random fashion, but rather simulating random behaviour. Importantly, the difference is that it is performing some computation tracking its immediate history so as not to repeat movements, given that probabilistically, statistical randomness would allow a long list of L or R, which is not optimal. Hence it has to go beyond probabilities and truly try to simulate algorithmic randomness, thus increasing the required computation to reproduce the desired random-looking behaviour and, therefore, its logical depth.

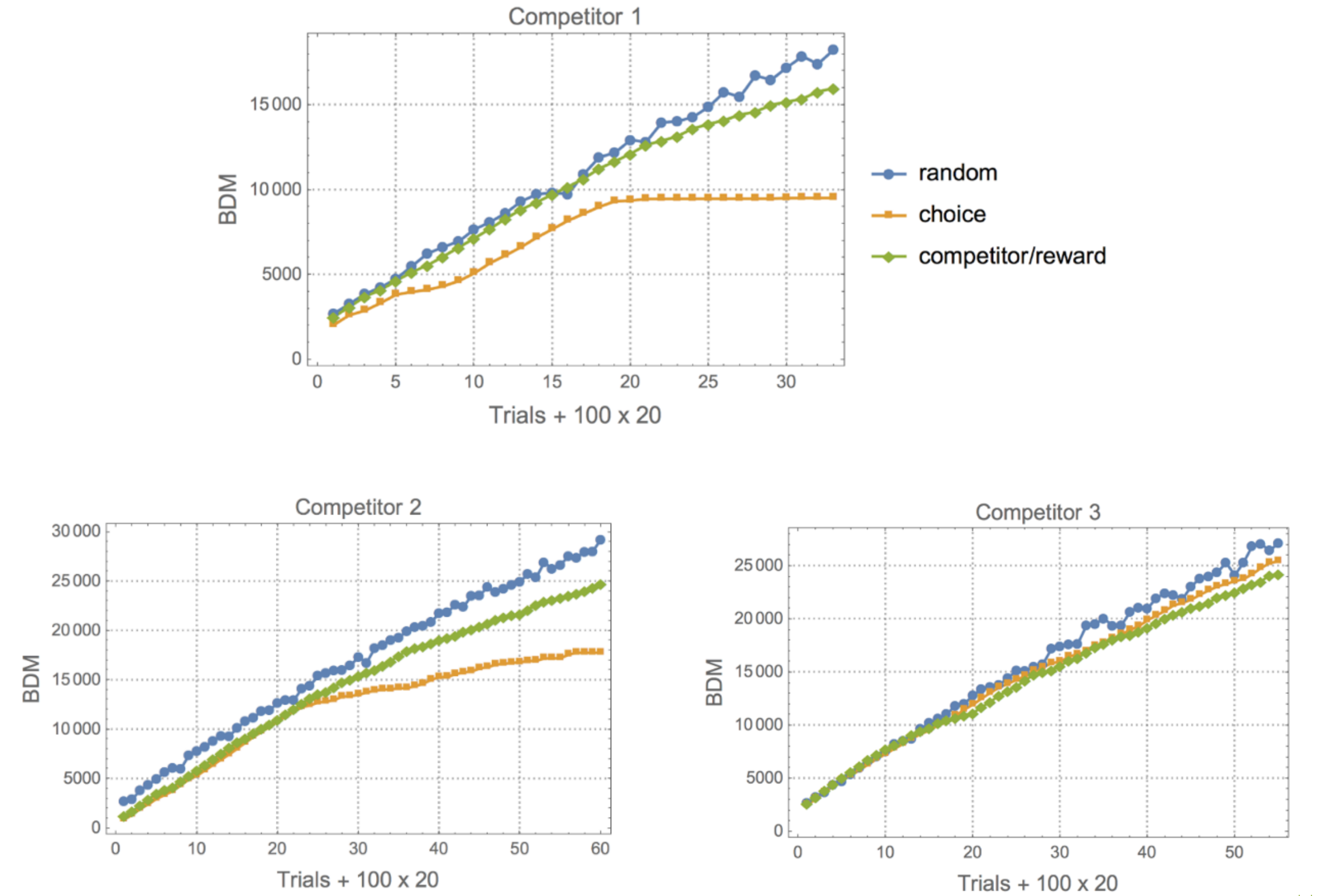

**Figure 4. Top: The same rat number 6 in the three different environments with different virtual competitors. With Competitor 1 this particular rat shows two kinds of "phase transitions," leading to a decrease in its behavioural complexity. The standard strategy appears to be to start off by displaying random behaviour to test the competitor's prediction capabilities, switching between different patterns up to a point where the rat settles on a successful strategy that allows it to simply repeat the same behaviour and yet receive the maximum reward after fooling the competitor. This phase comes later for Competitor 2 (Bottom left), given that this new virtual competitor is slightly more sophisticated. For Competitor 3, the rat is unable to outsmart it because it implements a more sophisticated predictive algorithm. The rat either cannot settle on a single strategy and keeps performing a random search or decides to switch to or remain in a particular mode after finding itself outsmarted by the virtual competitor.**

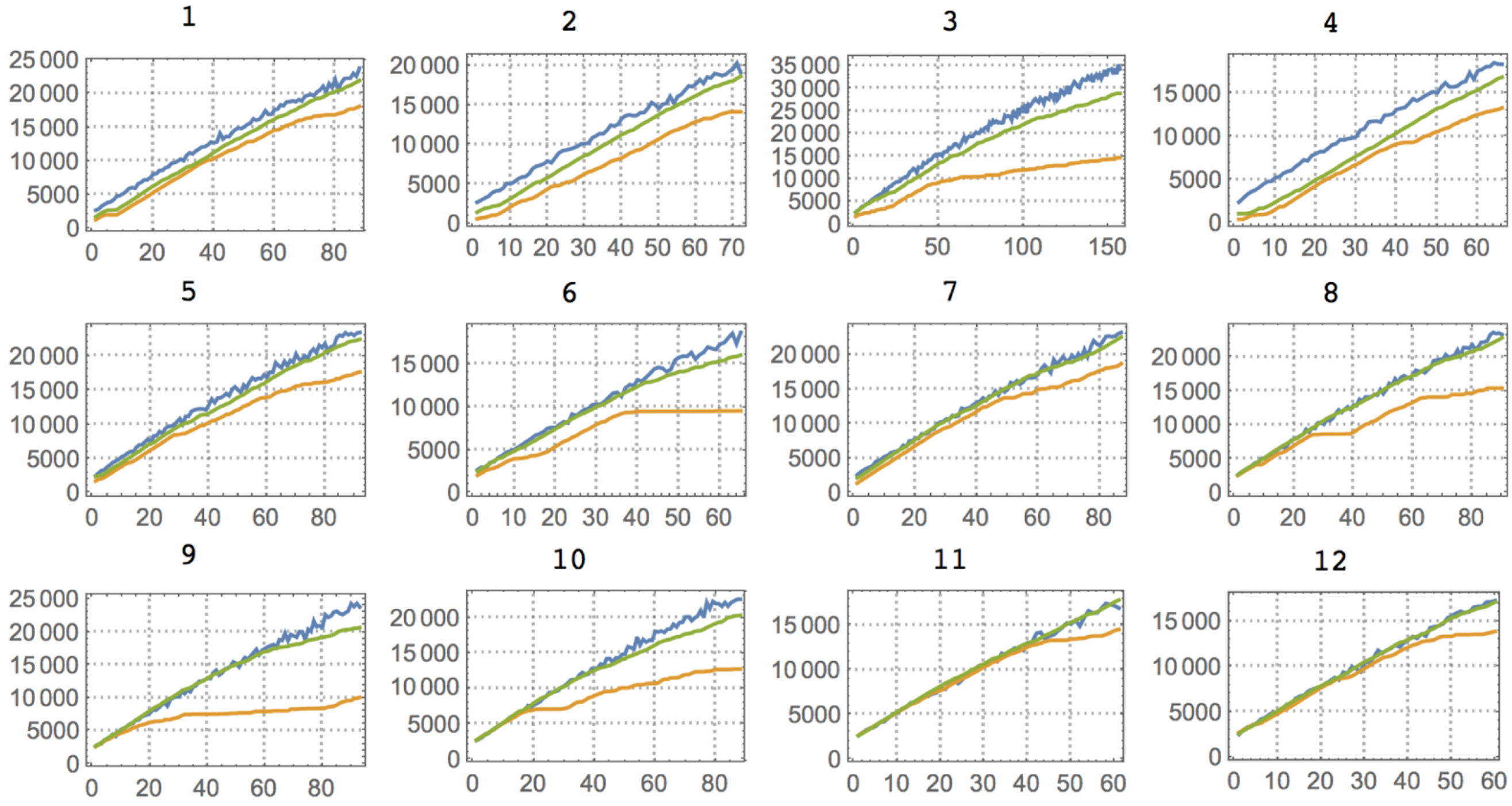

**Figure 5. Complexity (BDM) plots for all 12 individuals (rats) against Competitor 1. For Competitors 2 and 3, The most salient feature is that against Competitor 1 the rats show a quick decrease in complexity that allows them to keep the maximal reward (in all 3 cases. The rats were close to the maximal possible reward except for Competitor 3, where they continued to act randomly, either as a strategy or because they remained in the exploratory transition that is displayed at the beginning of every experiment). Colour legends and axis labels are the same as in Figure 5.**

Figures 4 and 5 show that these results suggest that the rats either switch to random behaviour on purpose or continue in the random mode they started as a testing strategy to gauge the competitor's capabilities. However, as shown by the clinical experiments, the rats seem to eventually suspend brain activity, seemingly after finding that they cannot devise an effective strategy (Tervo et al., 2014). But the results here build upon the previously reported conclusions that the rats seem to start with a high (random) complexity strategy in the first trials before settling on a single specific strategy if any. As shown in Figures 4 and 5, against Competitor 1, rats quickly decide on an optimal strategy of low complexity that keeps rewarding them, thanks to the poor predictive capabilities of the virtual competitor. But as Figure 4 shows a representative case (and not a special one for Competitors 2 and 3), we see that animals make the transition later (Competitor 2) or never make it (Competitor 3).

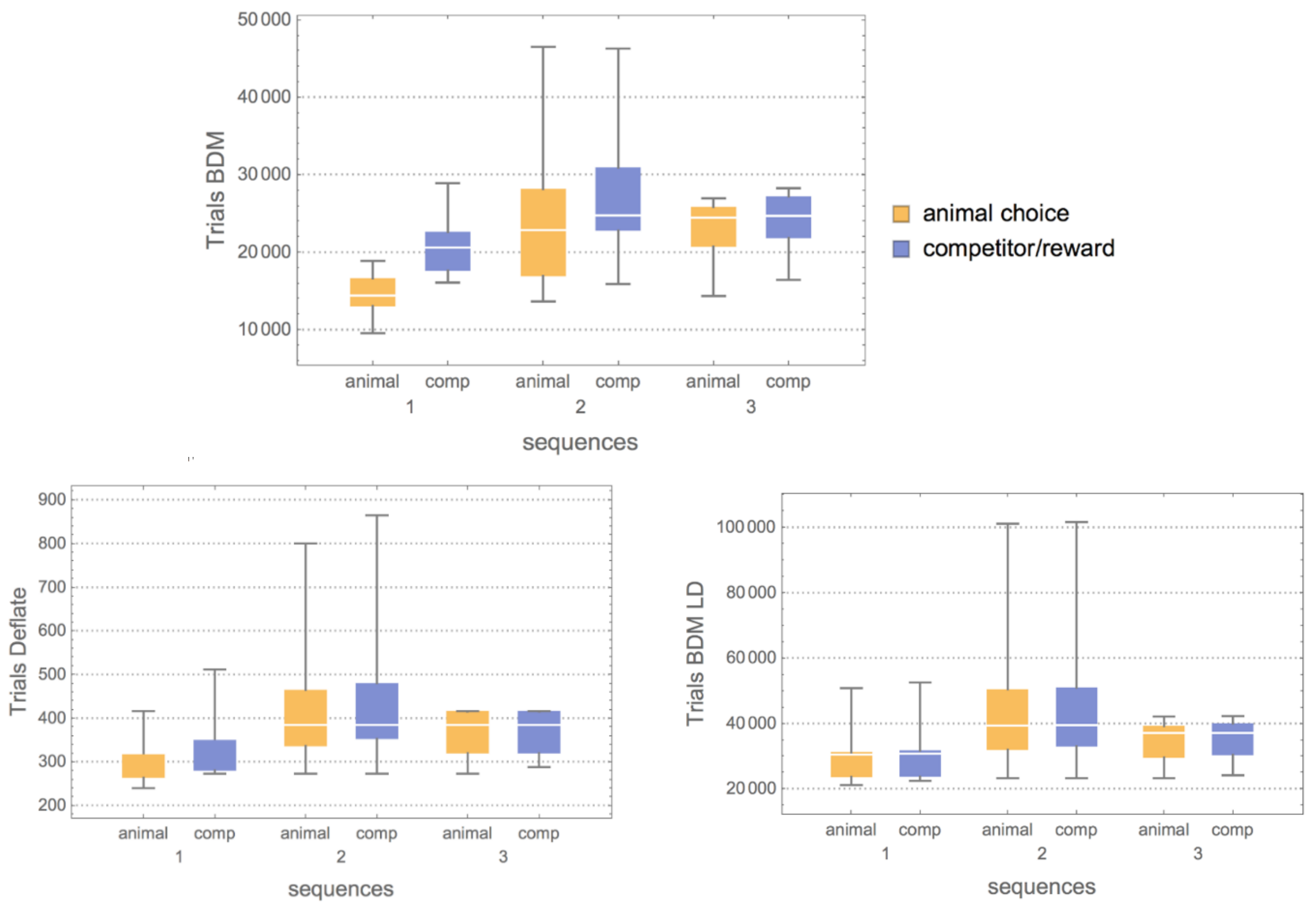

**Figure 6. Top: Box plots showing the results of three algorithmic information-theoretic complexity measures applied to the behavioural sequences for both animals and competitors (comp) in all three environments of increasing complexity (more powerful virtual competitor predictive capabilities). For BDM (Top), for Competitor 1, the complexity of both the animal and the competitor is low. Still, the gap between medians is extensive, meaning that the animal quickly outsmarted the competitor. For Competitor 2, however, not only is an increase in Kolmogorov complexity shown, but the median gap between the animal's behavioural complexity and the competitor's is smaller. Furthermore, the variance is greater, meaning that the animal explored more strategies of different complexity, and finally for Competitor 3 the medians match at higher Kolmogorov complexity (random-looking behaviour). Compress (implementing the Deflate lossless compression algorithm) confirmed the BDM results but showed less fine granularity than BDM allowed, likely due to the fact that lossless compression finds it difficult to detect small patterns that both the animals and these simple competitors rely so much upon. BDM LD confirmed the nature of the strategy against Competitor 3, where there is a decrease in structural complexity as compared to the interaction with Competitor 2, hence suggesting that the animal remained in a stochastic mode of low Logical Depth but high Kolmogorov complexity.**

The Spearman ranking tests among BDM and BDM LD in the three environments suggest, however, that intelligence can only be defined by task or environment in this experiment and for this set of animals. This is because no significant correlation was found in the ranking for different competitor experiments after a Kolmogorov-Smirnov test to determine the separation between animal and competitor behavioural complexity. There are two ways of outsmarting the virtual competitors that are confounded in the complexity curve. Either the animal learns fast and maximises gain (keeps the competitor's behaviour complex), or minimises effort (reflected in the decrease in its complexity). Hence the curves diverge.

**Table 3. A Kolmogorov-Smirnov comparison of the animal's choice and the competitor's complexity curves may provide a ranking of intelligence in a given environment, with the smallest values where the curves are at the greatest remove from each other and therefore signifying the fastest the animal has potentially settled on a successful strategy. No significant correlation was found among the different environments. Some animals consistently come out on top or at the bottom, which could be interpreted to mean that animals performed differently in different environments.**

| *Animal number* | *Kolmogorov-Smirnov statistic* |
|---|---|
| **3** | $4.88022 \times 10^{-60}$ |
| **6** | $1.083 \times 10^{-17}$ |
| **4** | $1.6514 \times 10^{-12}$ |
| **9** | $1.3695 \times 10^{-10}$ |
| **2** | $2.74921 \times 10^{-10}$ |
| **1** | $7.5875 \times 10^{-9}$ |
| **10** | 0.00045 |
| **8** | 0.001 |
| **5** | 0.003 |
| **7** | 0.012 |
| **11** | 0.223 |
| **12** | 0.55 |

Figure 6 shows how the structural complexity of the animal's behaviour matches that of the competitor, as suggested in Zenil et al., 2012b (this seems to indicate, however, that the match is with Logical Depth complexity and not Kolmogorov complexity). That estimation based upon or motivated by Logical Depth, a measure of sophistication, is greater in the second experiment means that indeed both the algorithm and the animal behaviour require more computational resources than in the 1st and 3rd experiments, where there is less consideration given to behavioural history, as the original paper reporting the clinical experiment claims. This also agrees with what we found by applying both lossless and BDM compression (Figure 4).

It is also interesting to look at the asymptotic behaviour of both the animal and the competitor (see Figure 5), as it indicates a period of training before the rats start overtaking the competitor with an optimally rewarding strategy. Indeed, for the experiment with Competitor 1, the training period is very short; on average between 100 and 300 trials are needed before the curves start to diverge, indicating that the animal has outsmarted the competitor. This can be advanced as a potential objective measure for animal intelligence, and one can see that subjects 3 and 6, for example, are among the fastest learners, the quickest at finding a good strategy, while subjects 11 and 12 are slow, a ranking based on a Kolmogorov-Smirnov test provided in Table 3. Against Competitor 2 (see Figure 4), the rat has a training period where it matches the virtual competitor's behaviour before outsmarting it. The gap indicates that there is a reward even in the face of lower animal complexity, which means it has cracked the competitor's behavioural code. Against Competitor 3 (see Figure 4), however, the animal is truly challenged and chooses to behave randomly, to which the learning algorithm reacts accordingly. Thus, the two match in manifesting high Kolmogorov complexity as compared to the previous cases.

For Competitor 2, the estimation motivated by Logical Depth increases because both the animal and the virtual competitor are engaged in a computation that requires slightly more computing power and time than when the animal is pitted against Competitor 1. However, for Competitor 3, the Logical Depth decreases, again as an indication of either greater simplicity or randomness, which in this case, agrees with the experiment. And taking into consideration the result with BDM, it is randomness that is introduced in the animal's choice of behaviour against Competitor 3, which is in full agreement with the results reported in the independent study (Tervo et al., 2014) and in accordance with the clinical experiments measuring cortical feedback to quantify brain activity during the performance of the tasks.

## Discussion and Conclusion

To recap, we first demonstrated the validity of our numerical approximation of algorithmic and structural complexity. These techniques are broadly useful as they provide the community with an objective tool to characterise "complexity" in behavioural experiments. Furthermore, the notion of structural complexity lets us glimpse the amount of computation that, generically, a system requires to perform a particular decision leading to a behavioural sequence.

Next, we assessed the applicability and the insights the notions of algorithmic and structural complexity could bring to well-known studies of animal behaviour involving ants, fruit flies, and rats. Importantly, these studies tax the animal to different degrees when it comes to environmental influence. One shared insight suggested by these studies, in light of our analysis, is that animals have some as yet unknown, mechanism(s) with which to perceive and cope appropriately with different degrees of complexity in their environment. In addition, beyond coping, it appears they can harness and utilise the environment in their internal decision process and in how they generate sequences of behaviour. For example, the less efficiently communicated instructions to ants have higher complexity, yielding longer communication time, resulting in more complex behavioural sequences. Here the K and LD metric provides a formal justification of the heuristic analysis performed by Reznikova et al in the original study (Ryabko and Reznikova, 1996). Hence, behavioural complexity matches environmental complexity. Furthermore, one common idea in the literature (e.g. Auger-Méthé, 2016) has been that animals navigating in an environment devoid of structural stimuli will adopt a random strategy. Our re-analysis, with formal complexity measures, of the fruit fly study suggests that flies are even more challenged to find different navigation strategies. The uniform group, having a uniform striped environment, was closed to an isotropic navigation strategy, closest to randomness or what is commonly referred to as Levy flight. In contrast, the group with a featureless environment was the most non-random group, high in logical depth, suggesting an algorithmic source for their decision process. Notably, this derived result aligns with the authors' (Maye et al., 2007) suggestion that the open-loop group appeared to be at the greatest remove from e randomness. Thus, in this case, a "simple" environment drives the animal toward an algorithmic bias in their attempt to devise a useful (navigation) strategy. Lastly, our analysis of the rats suggested a behavioural switch towards randomness in the face of environments with increasing complexity. Here, several interesting competitive situations were investigated. For example, our logical depth analysis revealed that the structural complexity of the rat always ends up matching the structural complexity of the competitor. In contrast, if the rat could not outsmart the competitor, it switched to random behaviour. Here the rat has to try to simulate algorithmic randomness to reproduce a random-looking behaviour. In the context of learning, deciding, and predicting, complexity measures captured subtle differences hinting at the mapping between an animal's environment, its sensory inputs, and its reactions. In all cases, we have seen animals react or adjust to the different scenarios, from communicating faster instructions to locate food to implementing deterministic and stochastic strategies against unknown environments and competitors. We have seen that animals can switch between a wide range of complex strategies, from behaving randomly against a competitor they cannot outsmart to behaving in a very structured fashion even in the absence of external stimuli, validating the results of Tervo et al. (2014).

In summary, we have shown that animal behavioural experiments can be analysed with novel and powerful tools drawn, based upon, or motivated by (algorithmic) information theory. Correlations in complexity can be established that are in agreement with and that elaborate on the conclusions of behavioural science researchers. We also report that a stronger correlation was consistently found between the behavioural sequences considered and their Kolmogorov complexity and Logical Depth than between these sequences and their Shannon Entropy. To the authors' knowledge, this is the first application of these tools to the field of animal behaviour and behavioural sequences.

Finally, we now turn to how these new results may translate to the understanding and analysis of human decision systems. Here we narrow the discussion on how to reinterpret experiments of how humans perceive randomness. Earlier pioneering studies by Kahneman D, Slovic P, Tversky A (1982) investigated how people reason and make decisions when confronted with uncertain and noisy information sources. Humans tend to report that a sequence of heads or tails, "HTTHTHHHTT", is more likely to appear than the series "HHHHHTTTTT" when a coin is tossed. However, the probability of each string is $1/2^{10}$, exactly the same, as indeed it is for all strings of the same length. In the "heuristics and bias" approach advocated and pioneered by Kahneman D, Slovic P, Tversky A (1982), these systematic errors were interpreted as biases inherent to human psychology or as the result of faulty heuristics. For instance, it was conjectured that people tended to say that "HHHHHTTTTT" was less random than "HTTHTHHHTT" because a so-called representativeness heuristic influenced them, according to which a string is more random the better it conforms to prototypical examples of random strings.

Given these purportedly faulty heuristics, human reasoning was interpreted as suggesting that humans had minds similar to a faulty computer. Considering recent work (Tervo et al., 2014), these behavioural biases can be accounted for by powerful complexity measures that point towards an algorithmic basis for behaviour. This suggestion accords with the results reported from animal studies (Ryabko and Reznikova 2009, Maye et al., 2007). In Zenil et al., (2012b) it was shown --- albeit in an oversimplified setting --- how animals would need to cope with environments of different complexity in order to survive and how this would require-- and possibly explain-- the evolution of information storage and the process of learning. In many ways, animal behaviour (notably, human behaviour) suggests that the brain often acts as a compression device. For instance, despite our very limited memory, we can retain long strings if they have low algorithmic complexity (Gauvrit et al., 2014). For the most part, cognitive and behavioural science deals with small sequences, often barely exceeding a few tens of values. For such short sequences, estimating the algorithmic complexity is a challenge. Indeed, prior to recent developments, the behavioural sciences relied largely on a subjective and intuitive account of complexity. While irreducibility is a pervasive challenge for the application of computation in data analysis (Zenil H et al., 2012d), the *Coding Theorem method* allows some experimental explorations based upon algorithmic complexity even on very short strings (Soler-Toscano et al., 2014) and objects in the context of an evolutionary process using the same tools (Hernández-Orozco et al. 2018). We suggest that these results align with our analysis of the animal experiments, that experiments on how humans perceive randomness suggests the existence of an algorithmic bias in our reasoning and decision processes. This contrasts with the view of the mind as performing faulty computations or random Levy flight computations when presented with items with some degree of randomness. It suggests that the brain has an algorithmic component. The "new paradigm" in cognitive science suggests that the human (and animal) mind is not a faulty machine but a probabilistic one, which estimates and constantly revises probabilities of events in the world, taking into account previous information and computing possible models of the future (Fahlman et al 1983; Rao and Ballard, 1999; Friston and Stephan, 2007; Friston, 2010).

While the construction of an internal mental model that effectively discerns the workings of a competitor could generate a successful counter-predictive strategy, apparently random behaviour might be favoured in situations in which the prediction of one's actions by a competitor or predator has adverse consequences (Nash, 1950; Maynard Smith and Harper, 1988). Notice how two of the experiments we considered in this paper, the rat experiment and that involving fruit flies, also relate to new developments in Integrated Information Theory (IIT), having to do with an approach to consciousness that proceeds via a mathematical formulation (Tononi, 2008). According to this theory, consciousness necessarily entails an internal experience. Here one indication of such an experience is the internal computation necessary to filter out or adopt non-random strategies in the absence of stimuli. Our results seem to support this view and provide further evidence of this hypothesis in line with IIT. Another is how the apparent randomness of a rat's behaviour may actually result in the rat engaging in a sophisticated computation, even while ignoring sensory input from the competitor or predator, as suggested in the clinical experiments. High Logical Depth indicates a causal history that requires more than simple feedforward calculations connecting sensors to actions. All this is also along the lines of more recent results where special predicting neurons, as distinct from mirror neurons, were found in monkeys. These neurons specialised in predicting an opponent's actions (Haroush and Williams, 2015).

In summary, the perspective brought to bear by our formal characterisation of complexity in the animal and human domain may, in turn, find application in designing cognitive strategies and measures in robotics and artificial intelligence (Zenil, 2015). The approach introduced here, based on algorithmic complexity measures, may help interpret data with tools complementary to the classical ones drawn from traditional statistics. For an animal to exploit the environmental deviation from equilibrium, animals must go beyond probabilities, i.e., beyond merely calculating the frequency of moves and beyond trivial entanglement with the environment. Animals clearly distinguish between environments of different complexity, reacting accordingly. The tools introduced here could contribute to modelling animal behaviour, discovering fundamental mechanisms, and also to the computational modelling of disease (Tegnér et al. 2009).